\documentclass[prb,aps,twocolumn,amsmath,amssymb,floatfix,superscriptaddress]{revtex4-2}
\usepackage{physics,graphicx,hyperref,mathtools}
\hypersetup{colorlinks=true,citecolor=blue,linkcolor=blue,urlcolor=blue}
\usepackage[utf8]{inputenc}
\DeclareUnicodeCharacter{2212}{-}
\DeclareUnicodeCharacter{03B1}{$\alpha$}
\DeclareUnicodeCharacter{03B2}{$\beta$}

\begin{document}

\title{Topologically nontrivial flat bands and quantum Hall crossovers in square-octagon lattice materials}

\author{Amrita Mukherjee}
\email{Contact author: amritaphy92@gmail.com}
\affiliation{Department of Condensed Matter Physics and Materials Science, Tata Institute of Fundamental Research, Mumbai 400005, India}

\author{Rahul Verma}
\affiliation{Department of Condensed Matter Physics and Materials Science, Tata Institute of Fundamental Research, Mumbai 400005, India}

\author{Pritesh Srivastava}
\affiliation{Department of Condensed Matter Physics and Materials Science, Tata Institute of Fundamental Research, Mumbai 400005, India}

\author{Bahadur Singh}
\email{Contact author: bahadur.singh@tifr.res.in}
\affiliation{Department of Condensed Matter Physics and Materials Science, Tata Institute of Fundamental Research, Mumbai 400005, India}

\begin{abstract}
Coexistence of nontrivial topology and flat electronic bands provides a fertile platform for correlated quantum states. The square-octagon lattice hosts Dirac nodes and flat bands at half-filling, yet the effects of intrinsic spin-orbit coupling (SOC) and staggered magnetic flux on its electronic and topological properties remain largely unexplored. Here, using tight-binding models incorporating SOC and staggered magnetic flux, we uncover a rich topological phase diagram in this lattice, comprising a quantum spin Hall phase with spin Chern number $C_s=1$, crossovers to quantum anomalous Hall phases with $C=1$ and $C=2$, and higher-order topological insulator phases with quantized quadrupolar corner charges. The initially dispersionless flat bands evolve into quasi-flat topological bands with nearly uniform quantum geometry and large flatness ratios, making them promising candidates for fractional Chern insulator states. We further identify realistic materials, including octagraphene, transition-metal dichalcogenides, synthetic $\mathrm{MoSi_2N_4}$, and magnetic $\alpha$-MnO$_2$, that may realize these tunable topological phases intertwined with flat-band physics, opening new opportunities for correlated topological matter.
\end{abstract}

\maketitle

\section{Introduction}
Flat-band materials with coexisting topological states are a cornerstone of condensed matter research. They reveal phases where topology, electronic correlations, and quantum geometry are deeply intertwined~\cite{sun2011nearly,jiang2019topological,ma2020spin,guan2023staggered,he2024topological}. 
Low-energy lattice models capture the essential electronic structure near the Fermi level while reducing materials-specific complexity to the most relevant degrees of freedom in such systems. They provide a transparent framework to understand band topology and emergent phases, as exemplified by the Haldane and Kane--Mele models~\cite{haldane1988model,kane2005quantum}.
Within this framework, lattice models hosting Dirac states and flat bands provide natural platforms for correlated topological phases, where spin--orbit coupling (SOC) or magnetic flux lifts band degeneracies, opens topological gaps, and reshapes the flat-band dispersion. Graphene, with a honeycomb lattice, serves as a prime example~\cite{geim2007rise}. It hosts Dirac points at the $K$ valleys that can be gapped to produce quantum spin Hall (QSH) and quantum anomalous Hall (QAH) states~\cite{kane2005quantum,kane2005z,haldane1988model,geim2007rise}. However, flat bands in graphene arise only through moiré superlattices, which require precise structural tuning and thus limit broader experimental accessibility~\cite{morell2010}. 

In contrast, Kagome and Lieb lattices naturally host flat bands due to quantum interference and geometric phase cancellation, avoiding complex structural control~\cite{ohgushi2000spin,lieb1989,leykam2018artificial}. Kagome metals such as AV$_{3}$Sb$_{5}$ (A = K, Rb, Cs) host diverse quantum phases, including nontrivial topology, time-reversal symmetry-breaking states, and superconductivity~\cite{lin2021complex}. Their Ti--Bi analogues, ATi$_{3}$Bi$_{5}$~\cite{yang2024superconductivity,patra2025high}, preserve kagome electronic features while exhibiting nematic and superconducting states. Lieb lattices, though rare in crystalline form, have been realized in molecular assemblies, photonic systems, cold atoms, and covalent organic frameworks, enabling controlled access to flat and Dirac bands in artificial lattices~\cite{slot2017experimental,mukherjee2015observation,taie2015coherent,cui2020realization}.

Square-octagon lattices have recently emerged as another promising two-dimensional geometry where Dirac states coexist with flat bands~\cite{kargarian2010topological,liu2013topological,nunes2020flat,wunderlich2023detecting}. Theoretical models predict two flat bands--one near half filling and another at the top of the spectrum--alongside van Hove singularities, which are points in the band structure where the density of states diverges. Staggered magnetic flux can induce nontrivial topology, producing QAH and higher-order topological insulator (HOTI) phases~\cite{pal2018nontrivial,he2022topological}. 
Staggered magnetic flux refers to alternating flux between neighboring plaquettes while maintaining zero net magnetic field within the unit cell, and has been widely used to explore topological phases and flat-band physics in lattice models~\cite{guan2023staggered,haldane1988model,he2024topological,chen2025}.
Material realizations of the square-octagon lattice include carbon allotropes such as octagraphene, transition-metal dichalcogenide monolayers, magnetic compounds including hollandites, and nitride monolayers~\cite{sheng2012octagraphene,liu2012structural,gorkan2019deformed,van2013grains,sun2015graphene,zhang2015two,
barik2021haeckelite,gurbuz2017single,verma2025emergent,lin2021ferromagnetism,crespo2013,PhysRevB.88.144428}. However, despite these theoretical and experimental advances, it remains unclear how intrinsic SOC and staggered flux influence the topological and flat-band characteristics of the square-octagon lattice. In particular, it is an open question whether these perturbations can generate nontrivial quantum geometry~\cite{roy2014band,ozawa2021relations,kruchkov2022} capable of supporting correlated phases such as fractional Chern insulators.

\begin{figure*}[t!]
\centering
\includegraphics[width=1.0\textwidth]{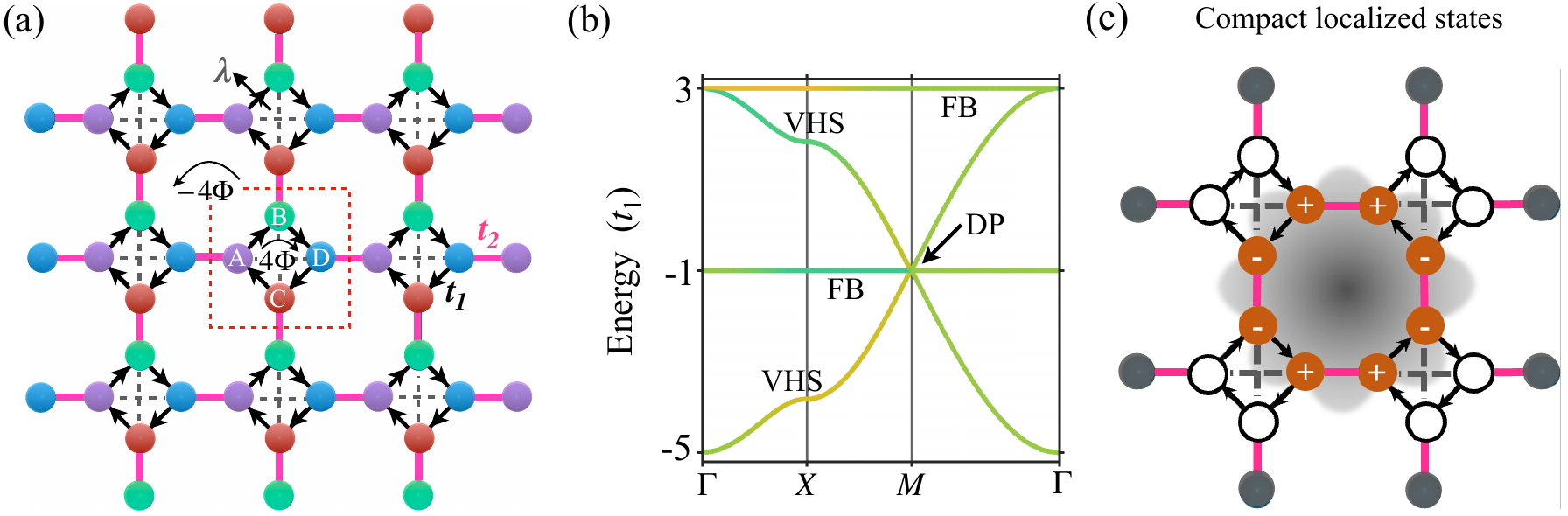}
\caption{Coexistence of flat bands and Dirac cones in a square-octagon lattice. (a) Square-octagon lattice with four sublattices (A-D) per unit cell, shown in distinct colors. $t_1$, $\lambda$, and $t_2$ represent the nearest-neighbor, next-nearest-neighbor, and intercell hopping parameters, respectively. Arrows indicate the magnetic flux directions. Square and octagonal plaquettes carry $4\Phi$ and $-4\Phi$ flux, respectively, yielding a staggered pattern with zero net flux per unit cell. (b) Energy dispersion for $t_2 = 2$ and $\lambda = t_1$. Van Hove singularities (VHSs), flat bands (FBs), and a Dirac point (DP) are marked. Colors denote the dominant sublattice contributions to each band. (c) Schematic of a compact localized state associated with the FB at $E =- t_1$ (gray). Filled and hollow sites correspond to nonzero and zero wavefunction amplitudes, respectively. The $\pm$ signs indicate phase alternation leading to destructive interference at the hollow sites, which localizes the state.}
\label{fig1}
\end{figure*}

Here we show that intrinsic SOC and staggered magnetic flux in the square--octagon lattice drive a series of topological crossovers among QSH, QAH, and HOTI phases. We trace the evolution of the flat bands and demonstrate how they acquire high flatness and nontrivial quantum geometry. Based on first-principles calculations, we identify realistic material candidates--including transition-metal dichalcogenides, $\mathrm{MoSi_2N_4}$ derivatives, and magnetic $\alpha$-MnO$_2$--as promising platforms for realizing switchable topological and flat-band phenomena. Our results establish the square-octagon lattice as a robust framework where nontrivial geometry and topology coexist to generate rich quantum states.

\section{Model Hamiltonian and band structure}
The square-octagon lattice comprises four sublattices per unit cell, as illustrated in Fig.~\ref{fig1}(a). It is characterized by nearest-neighbour (NN) intracell hopping $t_1$, intercell hopping $t_2$, and next-nearest-neighbour (NNN) diagonal hopping $\lambda$. For $t_2 = t_1$, the system forms an undimerized, uniform square-octagon network. A staggered magnetic flux applied to each square plaquette introduces an Aharonov-Bohm phase, such that NN hoppings acquire a factor $e^{\pm i\theta}$, where $\theta = \tfrac{2\pi \Phi}{4\Phi_0}$ and $\Phi_0 = \tfrac{hc}{e}$ is the magnetic flux quantum. The $\pm$ sign denotes the hopping direction along the square loop, as indicated by the arrows in Fig.~\ref{fig1}(a).  
The magnetic phase is assigned only to the $t_1$ hoppings, which gives a net phase $4\Phi$ around each square plaquette. Since the $t_2$ links that form the octagon remain flux-free, the total phase around an octagon is $-4\Phi$, which ensures zero net flux per unit cell and hence a staggered flux pattern. Such flux patterns can arise microscopically from loop-current states or orbital ordering, which generate complex hopping amplitudes and nontrivial Berry curvature in real materials (see Supplemental Materials (SM) for details~\cite{supplemental}).

In presence of staggered flux and intrinsic SOC, the system is described by a single-orbital tight-binding Hamiltonian,
\begin{equation}\label{ham}
    H = H_\mathrm{NN} + H_\mathrm{SOC}
\end{equation}
where $H_\mathrm{NN}$ and $H_\mathrm{SOC}$ denote the spin-independent hopping and intrinsic SOC terms, respectively:
\begin{equation}
H_\mathrm{NN} = -t_{1}\sum_{\langle j,k \rangle}  e^{i\theta} c_j^\dag c_k 
                                 -t_{2}\sum_{\langle j,k \rangle'} c_j^\dag c_k  
                                 -\lambda\sum_{\langle \langle j,k \rangle \rangle} c_j^\dag c_k 
                                 + H.c.,
    \end{equation}
and 
\begin{equation}
    H_\mathrm{SOC} = i\lambda_\mathrm{SOC}\sum_{\langle \langle j,k \rangle \rangle,\sigma \sigma^{'}} c_{j\sigma}^{\dagger}(\mathbf{e}_{jk}\cdot \boldsymbol{\sigma})c_{k\sigma^{'}}
    \end{equation}

Here, $\hat{c}_{j\sigma}^{\dagger}$ ($\hat{c}_{j\sigma}$) is the creation (annihilation) operator for an electron at site $j$ with spin $\sigma$. $\boldsymbol{\sigma}$ are the Pauli matrices, and $\lambda_\mathrm{SOC}$ is the intrinsic SOC strength. $\lambda_{\mathrm{SOC}}$ is expressed in unit of $t_1$. The factor $\mathbf{e}_{jk} = (\mathbf{d}_{jk}^{1} \times \mathbf{d}_{jk}^{2})_{z} = \pm 1$ encodes the direction of hopping between two NNN sites $j$ and $k$, defined by unit vectors $\mathbf{d}_{jk}^{1}$ and $\mathbf{d}_{jk}^{2}$ along the corresponding bonds~\cite{kane2005quantum,ma2020spin}.  We consider Kane–Mele-type intrinsic SOC with spin U(1) symmetry, which conserves spin and enables spin-resolved topology~\cite{kane2005quantum}. This is consistent with the materials studied here, where spin remains nearly aligned along the $z$-axis~\cite{verma2025emergent}.

\begin{figure*}[t!]
\centering
\includegraphics[width=0.92\textwidth]{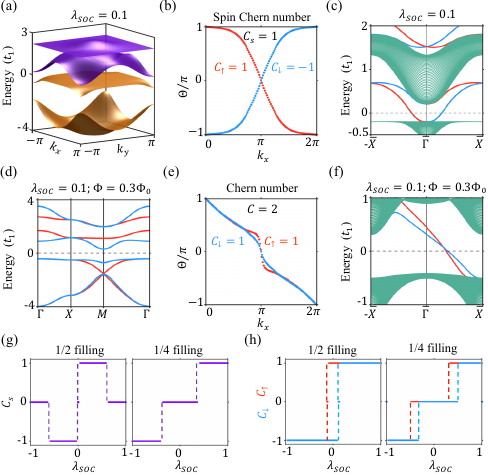}
\caption{Emergence of quantum Hall phases and nontrivial flat bands in square-octagonal lattice. (a) Band structure with SOC strength $\lambda_{\mathrm{SOC}}= 0.1$ and intercell hopping $t_2=1.2$. A nontrivial band gap opens across the full Brillouin zone (BZ) at half-filling, giving rise to a quantum spin Hall phase. Occupied and unoccupied bands are shown in orange and violet, respectively.  (b) Evolution of Wannier charge centers for two spin channels, exhibiting zero total Chern number but a finite spin Chern number.  (c) Edge-state spectrum along the (010) direction, showing spin-polarized edge modes. 
$\overline{\Gamma}$ and $\overline{X}$ denote high-symmetry points in the (010) edge-projected Brillouin zone.
(d) Band structure in the presence of SOC ($\lambda_{\mathrm{SOC}}=0.1$) and magnetic flux $\Phi=0.3\Phi_0$.  (e) Corresponding WCC evolution for two spin channels and (f) associated chiral edge states, indicating a quantum anomalous Hall phase with Chern number $C=2$ (QAH II).  (g) Evolution of the spin Chern number with SOC strength $\lambda_{\mathrm{SOC}}$ at 1/2 and 1/4 fillings.  (h) Evolution of Chern number as a function of magnetic flux $\Phi$ at the same fillings.}
\label{fig2}
\end{figure*}

\begin{figure*}[t!]
\centering
\includegraphics[width=1.0\textwidth]{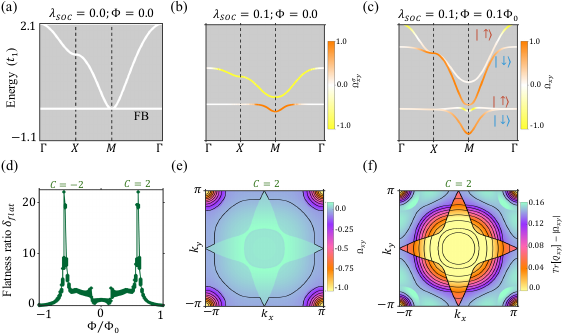}
\caption{Nontrivial flatness and quantum geometry. (a-c) Evolution of the flat band and Dirac cone at the $M$ point: (a) without SOC, (b) with SOC, and (c) with both SOC and magnetic flux. The gapless Dirac point in the absence of SOC evolves into a QSH phase with SOC and a QAH phase under magnetic flux.  Color maps indicate (b) the spin Berry curvature $\Omega_{xy}^{\sigma}$ and (c) the Berry curvature $\Omega_{xy}$ associated with band inversion. Flat bands acquire finite dispersion near the inversion point, forming quasi-flat topological bands.  (d) Flatness ratio of flat band (at half-filling) as a function of magnetic flux $\Phi$ for $t_2=1.2$ and $\lambda_{\mathrm{SOC}}=0.1$. Green regions denote the QAH phase with $C=\pm2$, where the topological flat band retains a high flatness ratio (see Ref.~\cite{supplemental} for energy dispersion). (e, f) Contour maps of (e) Berry curvature $\Omega_{xy}$ and (f) the difference between the trace of the quantum metric and the absolute value of Berry curvature, $\mathrm{Tr}[Q_{xy}] - |\Omega_{xy}|$, for the topological band with flatness ratio $\sim22$ and Chern numbers $C=2$ and $C=-2$.}
\label{fig3}
\end{figure*}

Under periodic boundary conditions, the Hamiltonian becomes
\begin{equation}
    H = \sum_{\vb*{k}} \psi^{\dag}_{\vb*{k}} \hat{\mathcal{H}}(\vb*{k}) \psi_{\vb*{k}}
\end{equation}
with spinor basis $\psi_{\vb*{k}}^{\dagger} = (\psi_{\vb*{k}\uparrow}^{\dagger}, \psi_{\vb*{k}\downarrow}^{\dagger})$, where $\psi_{\vb*{k}\sigma}^{\dagger} = (c_{A\vb*{k}\sigma}^{\dagger}, c_{B\vb*{k}\sigma}^{\dagger}, c_{C\vb*{k}\sigma}^{\dagger}, c_{D\vb*{k}\sigma}^{\dagger})$. $\hat{\mathcal{H}}(\vb*{k})$ represents the momentum-space Hamiltonian.
 Here, we set $t_1 = 1$ throughout our analysis, and all energies are measured in units of $t_1$.
 
For $\lambda = t_1$ and in the absence of SOC and magnetic flux, the eigenvalues of $\hat{\mathcal{H}}(\vb*{k})$ are
\begin{equation}
\left\{
\begin{split}
E_{n=1-4} &= t_{1} \pm t_{2},\\ 
&-t_{1} \pm \sqrt{4 t_{1}^2 +t_{2}^2 + 2t_{1}t_{2} (\cos{k_{x}}+\cos{k_{y}})}
   \end{split}
   \right.
   \label{sqoct_eigen}
\end{equation}
The corresponding band structure for $t_2=2$ and $\lambda=t_1$, with sublattice projections in the square Brillouin zone, is shown in Fig.~\ref{fig1}(b). The spectrum hosts two flat bands, one near half-filling and another at the top of the manifold. These flat bands touch dispersive bands at the $M$ point (half-filling) and the $\Gamma$ point ($3/4$ filling), forming symmetry-protected Dirac crossings. Notably, symmetry suppresses linear terms in the low-energy expansion near these points, leading to quadratic band touchings. While their energy positions shift with $t_2$, the flat bands remain nondispersive. Importantly, they are fully nondispersive across the Brillouin zone only for $\lambda = t_1$; for $\lambda \ne t_1$, they acquire dispersion and become partially flat~\cite{supplemental}.

The real-space localization of these flat bands is revealed by solving the Schrödinger equation $H \psi_{i} = E \psi_{i}$, where $\psi_{i}$ is the wavefunction at site $i$~\cite{maimaiti2017}. For the flat band at $E = -1$ [Fig.~\ref{fig1}(b)], the eigenstate exhibits complete localization over a finite cluster, with nonzero amplitudes confined within the cluster (brown sites) and vanishing amplitudes outside it (white sites), as illustrated in Fig.~\ref{fig1}(c). This compact localized state (CLS), analogous to those in Lieb and kagome lattices, underlies the nondispersive nature of the flat bands and their topological origin.

\section{Topological state crossovers}
The evolution of the square-octagon band structure with intrinsic SOC and magnetic flux $\Phi$ is shown in Fig.~\ref{fig2}. A finite SOC strength ($|\lambda_{\mathrm{SOC}}| > 0$) gaps the Dirac crossings, producing inverted band gaps at the $M$ point (half-filling) and the $\Gamma$ point ($3/4$ filling), as shown in Fig.~\ref{fig2}(a) for $\lambda_{\mathrm{SOC}} = 0.1$.  
We fit the model to first-principles band structures of realistic materials to set the parameter range~\cite{supplemental}. For square-octagonal MoS$_2$, $\lambda_{\mathrm{SOC}} \sim 0.02 eV$; here we use a slightly larger value to clearly illustrate the emergence of topological phases without losing generality~\cite{supplemental}.
Parity analysis yields $\mathbb{Z}_2 = 1$, confirming a topological insulating phase. This is further supported by the spin-resolved Wannier charge center (WCC) spectrum in Fig.~\ref{fig2}(b). The associated spin Chern number, obtained from the WCC flow, is $C_s = \tfrac{1}{2}(C_\uparrow - C_\downarrow) = 1$. Here, $C_\uparrow$ and $C_\downarrow$ denote the Chern numbers of the up- and down-spin channels, respectively. These are evaluated from the band-resolved Chern number,
\begin{equation}
C_{n} = \frac{1}{2\pi}\int_{BZ} d^{2}\vb*{k}~\Omega_{n,xy}(\vb*{k}),
\end{equation}
with the Berry curvature~\cite{xiao2010}
{\small
\begin{equation}
\Omega_{n,xy}(\vb*{k}) = \hbar^2\sum_{E_{m}\neq E_{n}}-2~\mathrm{Im}\frac{\langle \psi_{n}(\vb*{k})| v_x|\psi_{m}(\vb*{k})\rangle \langle \psi_{m}(\vb*{k})| v_y|\psi_{n}(\vb*{k})\rangle}{\left(E_{n}-E_{m}\right)^{2}},
\end{equation}}
where $|\psi_{n}(\vb*{k})\rangle$ is the $n^{th}$ Bloch state, and $v_x=\partial \hat{\mathcal{H}}(\vb*{k})/\partial k_{x}$, $v_y=\partial \hat{\mathcal{H}}(\vb*{k})/\partial k_{y}$ are velocity operators. From WCC analysis, $C_\uparrow = 1$ and $C_\downarrow = -1$, giving a total Chern number $C = 0$ (consistent with time-reversal symmetry) but a finite spin Chern number $C_s = 1$, characteristic of the QSH phase.

The edge spectrum in Fig.~\ref{fig2}(c) exhibits helical counterpropagating edge modes within the bulk gap, consistent with $\mathbb{Z}_2 = 1$ and $C_s = 1$. These helical states appear at both half- and $3/4$ filling, forming a ladder-like sequence of edge channels. As $\lambda_{\mathrm{SOC}}$ increases, the nontrivial gaps and edge states remain robust, while the flat bands evolve into quasi-flat bands.
The evolution of $C_s$ with $\lambda_{\mathrm{SOC}}$ at half- and quarter-filling, shown in Fig.~\ref{fig2}(g), exhibits a staircase behavior: $C_s = 1$ at small $\lambda_{\mathrm{SOC}}$ and drops to zero beyond a critical value ($\lambda_{\mathrm{SOC}} \gtrsim 0.6$), where the band gap closes and the system transitions to a gapless semimetal phase. The dependence of the topological gap on $\lambda_{\mathrm{SOC}}$ is shown in the SM~\cite{supplemental}. Reversing the sign of $\lambda_{\mathrm{SOC}}$ yields $C_s = -1$, reversing the edge-state chirality, analogous to the positive--negative SOC interplay observed in HgTe and HgSe~\cite{wang2015}. This tunability demonstrates that the QSH phase can be switched by locking or unlocking band inversions through SOC strength.

\begin{figure*}[ht!]
\includegraphics[width=1.0\textwidth]{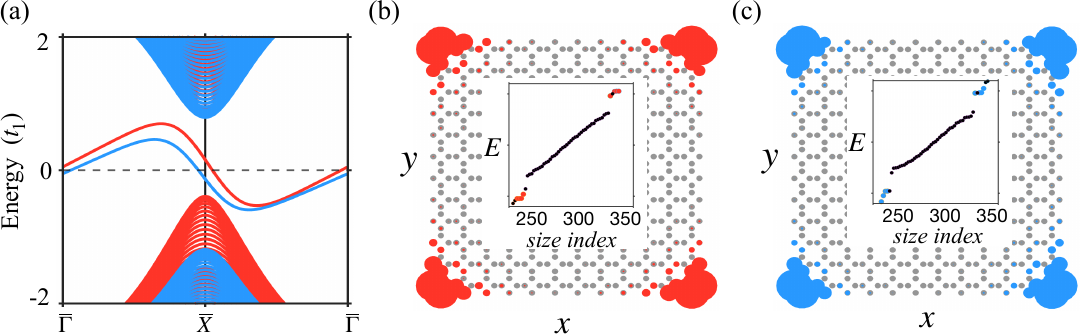}
\caption{Higher-order topological insulator with quadrupolar charge. (a) Edge band structure for magnetic flux $\Phi = 0.5\Phi_0$, intercell hopping integral $t_2=3.2$ and spin–orbit coupling $\lambda_{\mathrm{SOC}} = 0.1$, showing spin-polarized floating edge states. (b,c) Spatial distribution of corner-state wavefunctions $|\psi|^2$ at $E = t_1$ for up (red) and down (blue) spin channels. Fourfold-degenerate spin-polarized corner states appear at $E = \pm t_1$. Insets show the energy spectrum, where red and blue denote corner states and black denotes edge bands.}
\label{fig4}
\end{figure*}

Introducing a magnetic flux $\Phi$ breaks time-reversal symmetry, lifts Kramers degeneracy, and separates the spin channels, driving a crossover from the QSH to the QAH regime. For $\lambda_{\mathrm{SOC}} = 0.1$ and $\Phi = 0.3\Phi_0$, the band structure in Fig.~\ref{fig2}(d) shows lifted spin degeneracy, with unidirectional WCC flows for both spin channels (Fig.~\ref{fig2}(e)), resulting in a total Chern number $C = 2$. The corresponding edge spectrum in Fig.~\ref{fig2}(f) reveals two chiral edge states propagating in the same direction, defining a QAH-II phase ($C = 2$). With increasing $\Phi$, the inverted gap in one spin channel closes while widening in the other, leading to a QAH-I phase with $C = 1$. The evolution of spin-resolved Chern numbers with $\Phi$ (Fig.~\ref{fig2}(h)) shows that the QAH-II phase persists over a wide flux range at half-filling, while spin-selective QAH-I phases ($C = \pm1$) appear at quarter-filling. These results establish that both SOC and magnetic flux serve as effective tuning knobs for engineering topological crossovers in the square-octagon lattice.

\section{Emergence of nontrivial flat bands}  
We next examine how the flat bands evolve and acquire nontrivial topology under the combined effects of SOC ($\lambda_{\mathrm{SOC}}$) and magnetic flux ($\Phi$). When both $\lambda_{\mathrm{SOC}}$ and $\Phi$ are present, the perfectly flat band transforms into a quasi-flat band with a small dispersion near the $M$ point (Figs.~\ref{fig3}(a-c)). SOC induces finite spin-Berry curvature, while the magnetic flux breaks time-reversal symmetry and generates spin-dependent topological flat bands with nonzero Berry curvature. The flatness ratio, defined as $\delta_{\mathrm{flat}} = \Delta_{g}/\mathcal{W}$ (where $\Delta_{g}$ is the band gap and $\mathcal{W}$ the bandwidth)~\cite{guan2023staggered,he2024topological}, quantifies the propensity of a band to host interaction-driven topological phases. Its dependence on $\Phi$ and the intercell hopping $t_2$ for both spin channels is shown in Fig.~\ref{fig3}(d), along with the corresponding Chern numbers. Sharp peaks appear in narrow parameter ranges, with $\delta_{\mathrm{flat}}$ reaching values up to $\sim 22$ in the $C = \pm 2$ phase at half-filling for $t_2 = 1.2$ and $\Phi = 0.6\Phi_0$. 
The corresponding band structure with topological flat band is shown in the SM~\cite{supplemental}.

To assess the potential for fractional Chern insulators (FCIs), we further analyze the band geometry using the quantum geometric tensor~\cite{roy2014band,ozawa2021relations,kruchkov2022},
{\small
\begin{equation}
g_{n,xy}(\vb*{k}) = \sum_{E_{m}\neq E_{n}} \frac{\langle \psi_{n}(\vb*{k})| \tfrac{\partial \hat{\mathcal{H}}(\vb*{k})}{\partial k_{x}}|\psi_{m}(\vb*{k})\rangle \langle \psi_{m}(\vb*{k})| \tfrac{\partial \hat{\mathcal{H}}(\vb*{k})}{\partial k_{y}}|\psi_{n}(\vb*{k})\rangle}{(E_{n}-E_{m})^{2}},
\end{equation}}
where $|\psi_{n}(\vb*{k})\rangle$ is the $n$th Bloch state. The real and imaginary parts of $g_{n,xy}$ yield the quantum metric and Berry curvature, respectively: $Q_{xy} = \mathrm{Re}[g_{n,xy}]$ and $\Omega_{xy} = -2\mathrm{Im}[g_{n,xy}]$. The stability criterion for FCIs, $\mathrm{Tr}[Q_{xy}] - |\Omega_{xy}|$, quantifies the uniformity of the band geometry and topology across the Brillouin zone. In regions of high flatness, this difference remains nearly uniform (small but finite), indicating geometric conditions favorable for stabilizing FCIs. Two distinct regimes of enhanced flatness with $C = \pm 2$ are identified (Fig.~\ref{fig3}(d)). Figures~\ref{fig3}(e,f) present the Berry curvature $\Omega_{xy}$ and the corresponding $\mathrm{Tr}[Q_{xy}] - |\Omega_{xy}|$ distribution for the topological flat band with the optimal flatness ratio, confirming that the Chern band satisfies the geometric inequality dictated by the Fubini–Study metric~\cite{roy2014band,ozawa2021relations,kruchkov2022}. These results demonstrate that the square--octagon lattice, when tuned by SOC, magnetic flux, and hopping parameters, provides a robust platform with favorable band geometry for fractional topological phases.

\section{Higher-order topological state and phase diagram}   
Building on the discussion of QSH and QAH phases, we now examine how intercell hopping $t_2$ drives further topological transitions. For fixed $\lambda_{\mathrm{SOC}} = 0.1 t_1$ and $\Phi = 0.3\Phi_0$, increasing $t_2$ initially stabilizes a QAH I phase in the spin-up channel, characterized by a chiral edge state with $C = 1$. As $t_2$ increases, a subsequent inversion in the spin-down channel generates floating edge bands (FEBs) detached from the bulk continuum, while the spin-up channel retains $C = 1$. Upon further tuning $t_2$, both spin channels evolve into a HOTI with $C = 0$, where the FEBs remain isolated from the bulk, as shown in Fig.~\ref{fig4}(a) for $t_2 = 3.2$ and $\Phi = 0.5\Phi_0$.
This HOTI phase arises from strong intercell dimerization ($t_2 \gg t_1$), which induces band inversion and drives a transition from a first-order topological phase with edge states to a second-order phase with a quantized quadrupole moment. In this regime, the topology manifests as corner modes in a finite system, originating from mismatched edge dimerization and remaining well separated from the edge spectrum.

Although the total Chern number vanishes, the HOTI phase exhibits a quantized quadrupole moment $Q_{xy} = 1/2$. This quadrupolar moment is computed from the bulk polarization~\cite{benalcazar2017quantized},
\begin{equation}
\small
P_{x/y}^{\sigma} = -\frac{i}{2\pi} \log \Big[ \det \big(\nu_{k_{x/y}+2\pi \leftarrow k_{x/y}}^{\sigma}\big) \Big],
\end{equation}
where $\nu$ denotes the Wilson loop spectrum of the occupied bands for spin channel $\sigma$. The spin-resolved and total quadrupole moments are then given by
$
Q_{ij}^{\sigma} = P_{i}^{\sigma} P_{j}^{\sigma}, \quad i,j \in {x,y}, \quad
Q_{ij} = \sum_{\sigma} Q_{ij}^{\sigma}.
$
A finite $Q_{xy}$ ensures the presence of corner-localized modes coexisting with gapless FEBs. Figures~\ref{fig4}(b,c) display the energy spectrum of a nanodisk, where corner states for both spin channels appear within the bulk-edge gaps at discrete energies $E = \pm t_1$, exhibiting strong localization at the corners. This confirms that the HOTI phase in the square–octagon lattice hosts a robust quadrupolar corner charge, providing a platform where first- and higher-order topological phases coexist.

\begin{figure}[h!]
\includegraphics[width=0.5\textwidth]{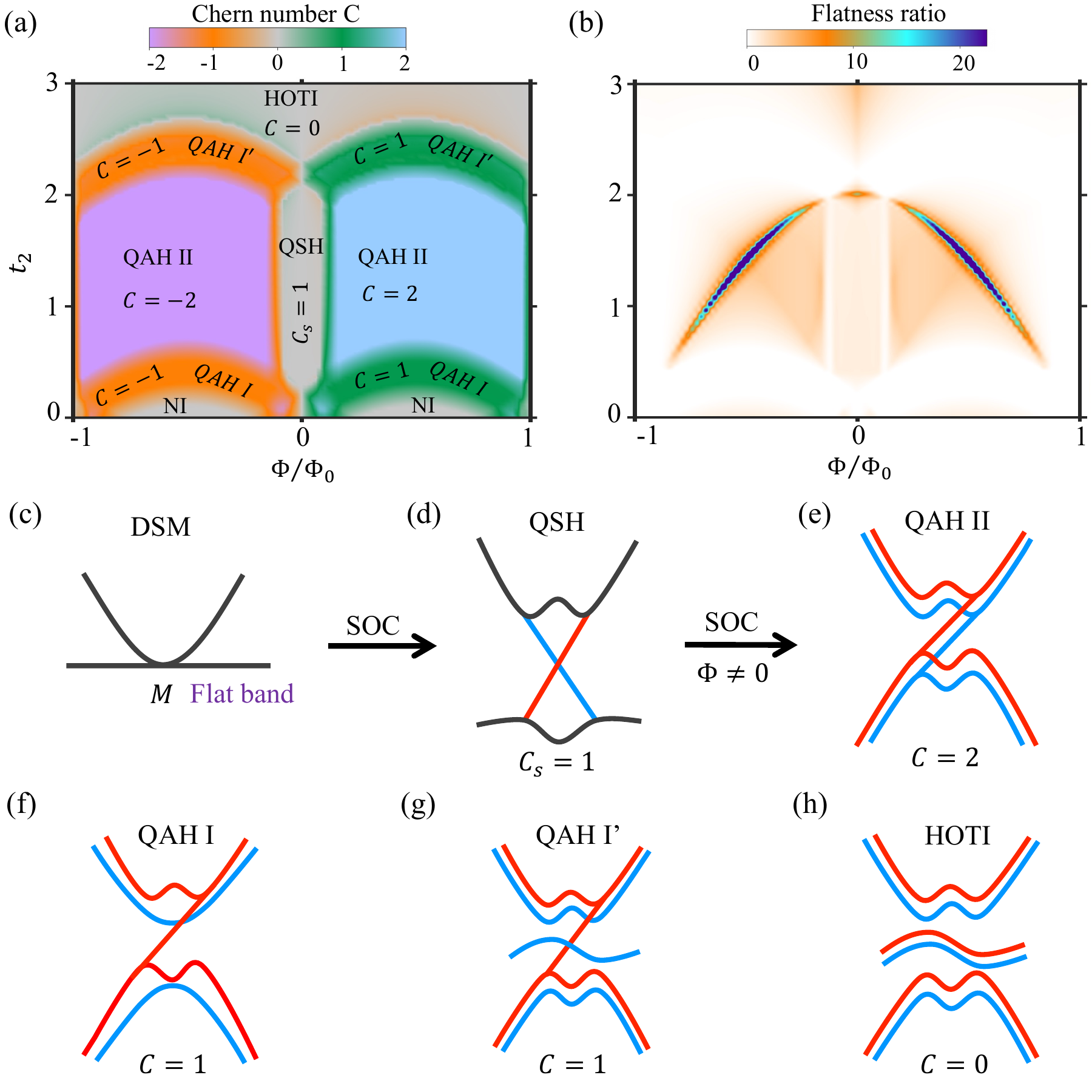}
\caption{ Topological phase diagram and spin-polarized topological flat bands. (a) Phase diagram with intercell hopping $t_2$ (vertical axis) and magnetic flux $\Phi$ (horizontal axis). The intrinsic SOC strength is fixed at $\lambda_{\mathrm{SOC}}=0.1$ with $\lambda=t_1$ and $t_1=1$. Distinct phases are identified by their Chern number $C$ or spin Chern number $C_s$: quantum anomalous Hall (QAH, QAH I, QAH I$^{\prime}$, QAH II), quantum spin Hall (QSH), higher-order topological insulator (HOTI), and normal insulator (NI). (b) Flatness ratio as a function of $t_2$ and $\Phi$ at half filling. (c-h) Representative band structures of the corresponding phases and the evolution of flat bands in the square-octagon lattice under SOC and magnetic flux. Red and blue denote spin-up and spin-down projections, while black lines represent spin-degenerate bands.}
\label{fig5}
\end{figure}

The complete topological phase diagram as a function of $t_2$ and $\Phi$ is shown in Fig.~\ref{fig5}(a). It reveals how the interplay between flux, SOC, and intercell hopping governs successive band inversions in the spin-resolved channels. For small $t_2$ and zero flux, SOC opens a nontrivial gap, realizing a QSH state with helical edge modes protected by time-reversal symmetry. Increasing $\Phi$ drives the system into the QAH I phase ($C = \pm 1$) with a single-spin inversion, and subsequently into QAH II ($C = 2$) with both spin band inversion that contribute co-propagating chiral edge states. At intermediate $t_2$, a second inversion in one spin channel gives rise to the QAH I$^{\prime}$ phase ($C = 1$) and generates a detached spin-polarized FEBs. Beyond a critical $t_2$, both spin channels undergo further inversion, stabilizing a HOTI phase ($C = 0$) with spin-degenerate corner states and FEBs. Specifically, for $t_2 \in [2.2, 2.6]$, one spin channel remains in the QAH state while the other develops FEBs, marking the transition region toward HOTI behavior. Such FEBs, rarely reported in prior studies, can serve as distinct hallmarks of higher-order topology in the square-octagon lattice. Moreover, the topological flat bands evolve into quasi-flat bands with enhanced flatness ratios, as shown in Fig.~\ref{fig5}(b). Figures~\ref{fig5}(c-h) schematically summarize the corresponding band structures across these topological regimes.

\begin{figure*}[ht!]
\includegraphics[width=1\textwidth]{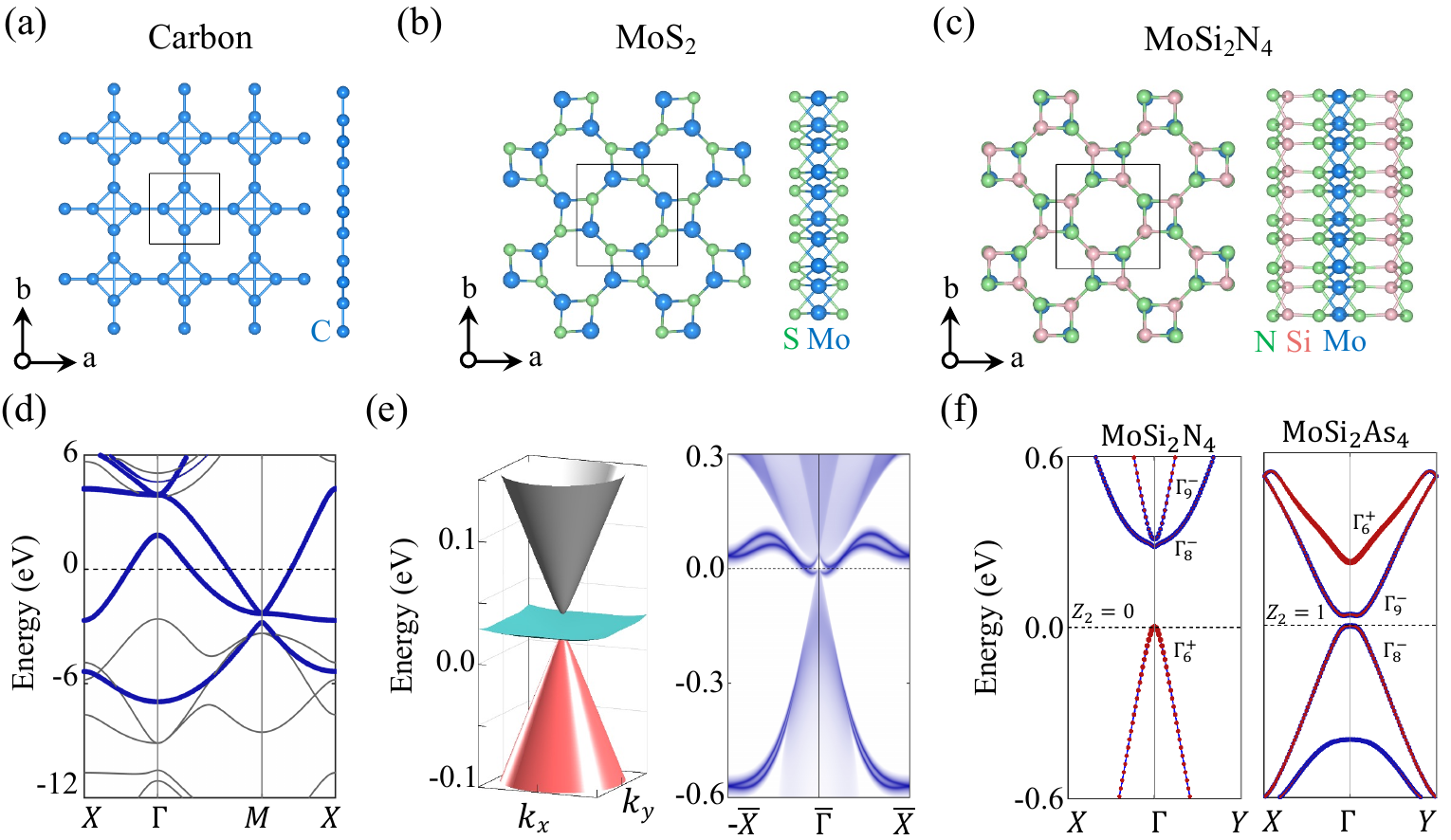}
\caption{ Materials realizing square–octagonal lattice geometry: Top and side views of square-octagon monolayers of (a) carbon, (b) transition metal dichalcogenide MoS$_2$, (c) bottom-up designed synthetic MoSi$_2$N$_4$. The corresponding electronic band structures are shown for (d) carbon, (e) MoS$_2$, (f) MoSi$_2$N$_4$. These systems exhibit diverse quantum phases, including a metallic state in carbon, coexisting Dirac-like bands and a quantum spin Hall state with a van Hove singularity near the Fermi level in MoS$_2$, a tunable topological state in MoSi$_2$N$_4$.}
\label{fig6}
\end{figure*}

\section{Materials realization}
We explored possible materials hosting the square-octagonal lattice geometry. This lattice appears across several material classes, including carbon allotropes, group-Va elements, transition-metal dichalcogenides (TMDs), nitrides, phosphides, carbides such as GaN and AlN, the synthetic MoSi$_2$N$_4$ family, and magnetic compounds including MnN and hollandites~\cite{sheng2012octagraphene,liu2012structural,gorkan2019deformed,van2013grains,sun2015graphene,zhang2015two,
barik2021haeckelite,gurbuz2017single,verma2025emergent,lin2021ferromagnetism,crespo2013,PhysRevB.88.144428}. Here, we focus on a few representative examples, shown in Fig.~\ref{fig6}, which illustrate the diversity of tunable electronic and topological phases in nonmagnetic systems.

First-principles calculations were performed within the density functional theory (DFT) framework using projector augmented-wave (PAW) potentials as implemented in the Vienna \textit{ab initio} simulation package (VASP)~\cite{hohenberg1964,kreese1996}. Exchange-correlation effects were treated using the generalized gradient approximation (GGA), and SOC was included self-consistently to account for relativistic effects~\cite{perdew1996}. Material-specific tight-binding Hamiltonians were constructed via the \textsc{VASP2WANNIER90} interface~\cite{mostofi2008wannier90}, and edge states were obtained using the semi-infinite Green’s function method~\cite{sancho1985highly}. Further details of the lattice and computational parameters are provided in the SM~\cite{supplemental}.

Figure~\ref{fig6}(a) shows the crystal structure of octagraphene (T-graphene), where carbon atoms form a square-octagonal lattice with unequal $sp^2$ hybridization arising from distinct inter- and intra-square C-C bond lengths. The orbital-resolved band structure (Fig.~\ref{fig6}(d)) reveals metallic behavior, with states near the Fermi level dominated by carbon $p_z$ orbitals (blue). The dispersion exhibits electron and hole pockets at the $\Gamma$ and $M$ points, respectively, along with quasi-flat bands along $\Gamma-X$ and $M-X$ directions.

In TMDs, the square-octagonal geometry can form through $4$-$8$ defects at grain boundaries of monolayer MoS$2$~\cite{sun2015graphene,van2013grains}. Structurally, these square--octagonal motifs emerge at grain boundaries, where Mo atoms remain coordinated by six S atoms. The calculated electronic dispersion near the Fermi level without SOC (Fig.~\ref{fig6}(e)) shows semimetallic behavior, featuring Dirac-like bands and a nearly flat band at $\Gamma$. The top valence and bottom conduction bands, dominated by Mo $d_{z^2}$ and $d_{x^2-y^2}$ orbitals with opposite parities, are degenerate at the Fermi level. SOC lifts this degeneracy, opening a gap and producing a QSH phase with nontrivial $\mathbb{Z}_2=1$ and $C_s=1$. The corresponding helical edge states are shown in Fig.~\ref{fig6}(e). Breaking $C_4$ symmetry splits the $\Gamma$-point Dirac node into two Dirac cones, leading to van Hove singularities at $\Gamma$~\cite{sun2015graphene}. Strain further drives topological phase transitions, tuning MoS$_2$ between QSH, trivial insulating, and metallic phases.

Synthetic MA$_2$Z$_4$ ($M =$ Mo/W, $A =$ Si/Ge, $Z =$ pnictogen) monolayers with a square--octagonal lattice offer another tunable platform for topological phases and flat bands. The square--octagonal polymorphs are predicted as stable variants of experimentally realized 2H-MoSi$_2$N$_4$~\cite{science2020}. Their stability is supported by total-energy calculations, phonon spectra without imaginary modes, and \textit{ab initio} molecular dynamics at 300 K~\cite{verma2025emergent}. Energy comparisons with other phases (1H, 1T, and 1T$'$) place the square--octagonal phase in a metastable yet experimentally accessible regime, similar to TMDs.
Their central MoN$_2$ layer resembles that of TMDs and is sandwiched between two SiN layers, forming a seven-atom-thick structure. Figure~\ref{fig6}(f) displays the band structures and irreducible representations near the Fermi level for two representative compositions. In MoSi$_2$N$_4$, the valence band maximum at $\Gamma$ primarily derives from Mo $d_{x^2-y^2}$ orbitals (red), while the conduction band minimum originates from doubly degenerate $d_{z^2}$ orbitals (blue), yielding a topologically trivial direct-gap insulator ($\mathbb{Z}_2=0$). Substituting N with As induces a band inversion at $\Gamma$, producing a doubly degenerate state at the Fermi level. SOC then opens a gap, realizing a QSH phase with $\mathbb{Z}_2=1$ and $C_s=1$. Consequently, MoSi$_2$(N$_x$As$_{1-x}$)$_4$ enables continuous tuning between trivial and topological insulating states via chemical substitutions.

Finally, several magnetic compounds, such as hollandites~\cite{crespo2013,PhysRevB.88.144428}, also crystallize in square-octagonal frameworks and can host spin-polarized topological states. Their double chains of edge-sharing MnO$_6$ octahedra form distorted square-octagon motifs, generating partially flat $d_{xy}$-derived bands and gapped Dirac crossings at $\Gamma$ and $M$ as shown for $\alpha$-MnO$_2$ in the SM~\cite{supplemental}. The resulting magnetic frustration and anisotropy lead to competing magnetic orders, offering a possible route to spin-liquid-like behavior.

\section{Summary}
We have demonstrated a rich landscape of topological phases in a two-dimensional square-octagon lattice that hosts both Dirac states and nontrivial flat bands. This lattice supports two flat bands: one pinned at the apex of the Dirac cone at half-filling and the other capping the overall band dispersion. The inclusion of SOC opens nontrivial energy gaps at the Dirac points, realizing a QSH phase with $C_s = 1$. Introducing a staggered magnetic flux $\Phi$ breaks time-reversal symmetry, lifts the Kramers degeneracy, and induces QAH phases with Chern numbers $C = 1$ and $C = 2$. Systematic tuning of the intercell hopping parameter $t_2$, which governs lattice dimerization, drives topological crossovers from QAH phases with chiral edge modes to HOTI phases characterized by floating edge bands and a quantized quadrupolar charge. The interplay among SOC, magnetic flux, and $t_2$ further renders the flat bands quasi-flat and topologically nontrivial, with regions of parameter space exhibiting remarkably high flatness ratios of $\sim 22$ for $C = \pm 2$. These quasi-flat bands possess nearly uniform quantum geometric tensor distributions, establishing favorable conditions for realizing fractional Chern insulator states.

We have also identified realistic material candidates with square–octagonal lattice geometry, such as octagraphene, MoS$_2$, and MoSi$_2$(N, As)$_4$, where Dirac-like and flat-band features dominate the low-energy electronic structure, facilitating diverse topological phases. Unlike in kagome or Lieb lattices, the flat bands in the square-octagon system are intrinsically pinned at the Dirac point at half-filling, making them more robust and conducive to interaction-driven topological phenomena. Our results establish the square–octagon lattice as a versatile and tunable platform for engineering coexisting topological and flat-band phases, opening promising avenues for exploring correlated topological matter and potential applications in next-generation electronic and spintronic devices.

\section{Acknowledgement}
We thank Bikash Patra for helpful discussions. We acknowledge the support of the Department of Atomic Energy, Government of India, under Project Identification Nos. RTI4013 and RTI4015. 

\nocite{dudarev1998electron,perdew1996,wu2018wanniertools5,mostofi2008wannier90,bounoua2020loop,Ye2015,Huang2017,sancho1985highly,hohenberg1964,kreese1996,perdew1996,crespo2013,haldane1988model}
\bibliography{ref}

\end{document}


\title
{
{-- Supplemental Material --\\
Topologically nontrivial flat bands and quantum Hall crossovers in square-octagon lattice materials}
}

\author{Amrita Mukherjee}
\email{Contact author: amritaphy92@gmail.com}
\affiliation{Department of Condensed Matter Physics and Materials Science, Tata Institute of Fundamental Research, Mumbai 400005, India}

\author{Rahul Verma}
\affiliation{Department of Condensed Matter Physics and Materials Science, Tata Institute of Fundamental Research, Mumbai 400005, India}

\author{Pritesh Srivastava}
\affiliation{Department of Condensed Matter Physics and Materials Science, Tata Institute of Fundamental Research, Mumbai 400005, India}

\author{Bahadur Singh}
\email{Contact author: bahadur.singh@tifr.res.in}
\affiliation{Department of Condensed Matter Physics and Materials Science, Tata Institute of Fundamental Research, Mumbai 400005, India}

\date{\today}
\maketitle

\subsection{Tight-binding model parameters}
\begin{figure}[h!]
\includegraphics[width=0.5\textwidth]{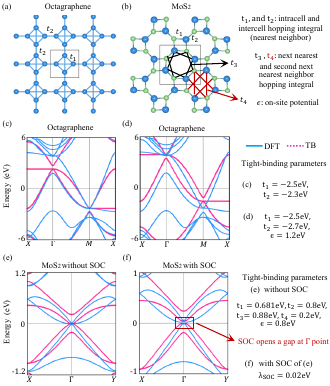}
\caption{(a,b) Crystal structures of octagraphene and MoS$_2$. Calculated band structures of (c,d) octagraphene and (e,f) MoS$_2$ obtained from DFT (blue solid lines) and the tight-binding model (pink dashed lines). Panels (c, d, e) show results without SOC, while (f) include SOC. Fitting parameters are also provided.}
\label{fig:S1}
\end{figure}

To obtain realistic tight-binding (TB) parameters used in the main text, we fit our calculated first-principles band structures of octagraphene and square-octagonal MoS$_2$. Figures~\ref{fig:S1}(a,b) illustrate the TB lattice models for octagraphene and MoS$_2$, respectively, along with the fitted energy dispersions obtained from the tight-binding model given in Eq.~1 of the main text. For MoS$_2$ (Figs.~\ref{fig:S1}(e,f)), the TB fitting reproduces the main features of the first-principles band structure, including Dirac-like states at the $\Gamma$ point and degeneracies at the $X (Y)$ point. When spin--orbit coupling (SOC) is included, a band gap opens at the $\Gamma$ point, consistent with density functional theory (DFT) results. From the fitting, the intrinsic SOC strength is estimated to be $\lambda_{\mathrm{SOC}} \approx (0.02$--$0.03) eV$. Although SOC is relatively small in these systems, its magnitude depends on the atomic species and local symmetry and can increase in materials containing heavier elements. To clearly resolve the topological gaps and phases, we therefore adopt a somewhat larger value, $\lambda_{\mathrm{SOC}} \approx 0.1t_1$, in the model with $t_1=1$ and energies are measured in units of $t_1$.

The intercell hopping parameter $t_2$ originates from bonds whose lengths are comparable to those associated with the nearest-neighbor hopping $t_1$, leading to $t_2 \sim t_1$, with typical deviations $|t_1 - t_2| \lesssim 0.2 eV$. For octagraphene (Figs.~\ref{fig:S1}(c,d)), the TB fitting captures the essential low-energy features near the Fermi level, including Dirac points and flat bands. Using a similar hopping pattern as in MoS$_2$, we obtain good agreement with the first-principles band structure, supporting the realistic parameter range of $t_2$ and $\lambda_{\mathrm{SOC}}$ used in our model.

\subsection{Variation of topological band gap with SOC strength}
\label{topological band}

\begin{figure}[t!]
\includegraphics[width=0.45\textwidth]{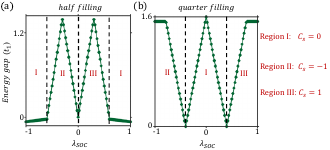}
\caption{Topological band gap as a function of the spin--orbit coupling $\lambda_{\mathrm{SOC}}$ at half- and quarter-fillings. The corresponding phases are distinguished by the spin Chern number $C_s$.}
\label{fig:S2}
\end{figure}

To demonstrate the effect of SOC, we plot the band gap as a function of $\lambda_{\mathrm{SOC}}$ for the nearest-neighbor intercell hopping $t_2 =1.2$ at half- and quarter-fillings (Fig.~\ref{fig:S2}). At half-filling, the gap initially increases from a nearly gapless state, reaches a maximum, and then gradually decreases as the SOC strength approaches $t_1$. In contrast, at quarter filling, the gap decreases, closes at a critical value, and subsequently reopens with increasing SOC, indicating a topological phase transition. The corresponding topological phases are summarized in Fig.~2(g) of the main text. To further characterize these phases, we compute the spin Chern number $C_s$, whose change between finite and zero values reflects the associated band inversion.
The topological nature of each regime is characterized by the spin Chern number $C_s$. A nonzero value of $C_s$ corresponds to a quantum spin Hall phase supporting spin-polarized edge states, whereas $C_s = 0$ indicates a trivial insulating phase.

\subsection{Condition for the emergence of a flat band}
\label{diagonal hopping}
\begin{figure}[tb!]
\includegraphics[width=0.45\textwidth]{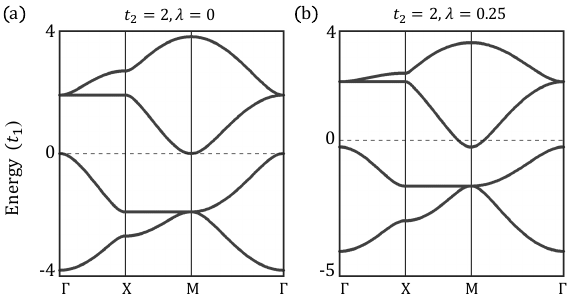}
\caption{Evolution of the electronic band structure with increasing next-nearest-neighbor diagonal hopping parameter $\lambda$, demonstrating the absence of a completely flat band. (a) $\lambda=0$, (b) $\lambda=0.25$.}
\label{fig:S3}
\end{figure}

In our tight-binding model, when the next-nearest-neighbor diagonal hopping integral $\lambda$ differs from $t_1$, the bands remain weakly dispersive and exhibit partially flat features only along certain high-symmetry directions in the Brillouin zone. This behavior is illustrated in Figs.~\ref{fig:S3}(a) for $\lambda = 0$ and \ref{fig:S3}(b) for $\lambda \neq 0$.

A fully flat band emerges only when the condition $\lambda = t_1$ is satisfied, which suppresses dispersion along the relevant directions and produces a non-dispersive band. Figures~\ref{fig:S3}(a,b) show the band dispersions for $\lambda = 0$ and $\lambda = 0.25$, where flat bands appear only along specific directions ($\Gamma$--X for the conduction band and X--M for the valence band).

\begin{figure}[t!]
\includegraphics[width=0.48\textwidth]{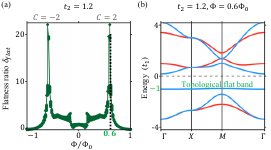}
\caption{ (a) Flatness ratio as a function of magnetic flux $\Phi$ for $t_2=1.2$. The topological flat band with $C=2$ attains its maximum flatness ratio ($\approx 22$) at $\Phi=0.6\Phi_0$ (green dashed line). (b) Band dispersion for $t_2=1.2$ and $\Phi=0.6\Phi_0$, highlighting the topological flat band at $E=-t_1$ (green). The flat bands for both spin projections appear at nearly same energy level. }
\label{fig:S4}
\end{figure}

\subsection{Topological flat band with high flatness ratio} \label{flatness}
We analyze the topological flat band with the highest flatness ratio ($\approx 22$). Figure~\ref{fig:S4}(a) shows the flatness ratio as a function of magnetic flux $\Phi$, revealing the optimal value $\Phi = 0.6\Phi_0$ (green dashed line). The corresponding band dispersion is shown in Fig.~\ref{fig:S4}(b), where the topological flat band appears just below the Fermi level at $E=-t_1$ (green). The flat bands for both spin projections lie nearly at the same energy below the Fermi level at $E=-t_1$. Their associated results are shown in Fig.~3(e, f) of the main text. 

\subsection{Staggered magnetic flux and real materials} \label{QAH_square-octagon}
\begin{figure*}[t!]
\includegraphics[width=\textwidth]{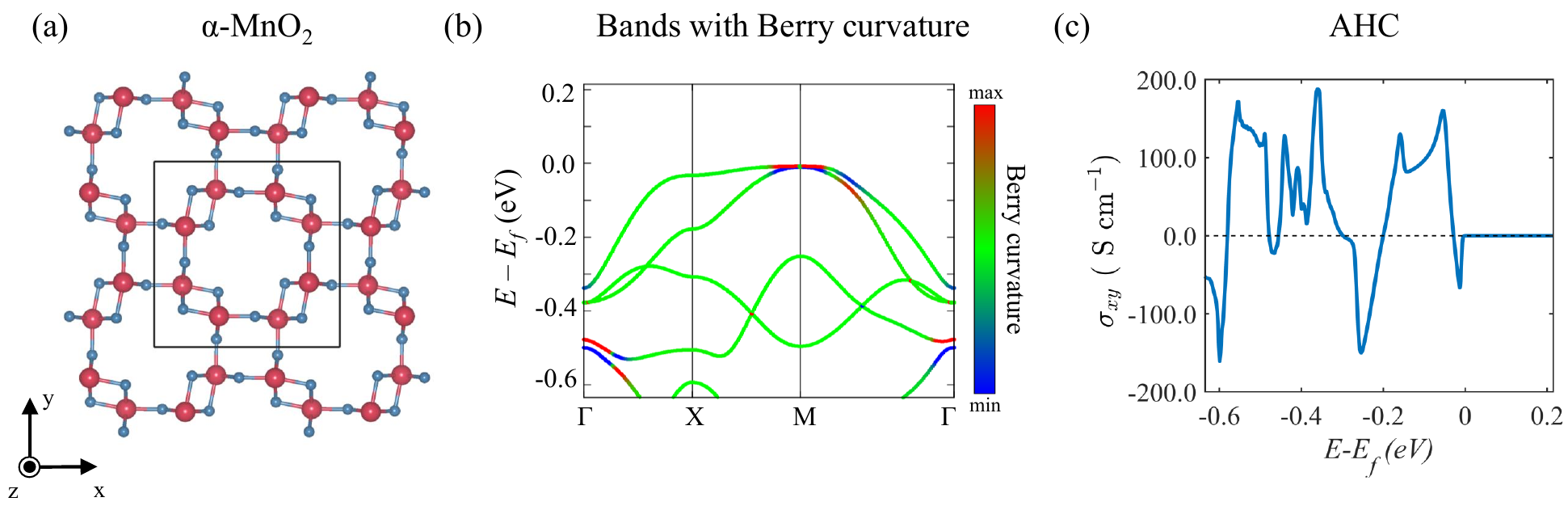}
\caption{(a) Top view of the crystal structure of $\alpha$-MnO$_2$. Red (blue) balls represent Mn (O) atoms. (b) DFT-calculated band structure and Berry curvature of the ferromagnetic phase with spin polarized along $z-$ axis using GGA+$U$ with $U = 3$ eV. (c) Corresponding anomalous Hall conductivity $\sigma_{xy}$, showing the quantum anomalous Hall effect.}
\label{fig:S5}
\end{figure*}

Our tight-binding model identifies the minimal symmetry conditions for a staggered-flux-driven quantum anomalous Hall effect (QAHE) in square–octagon lattices, analogous to the Haldane model~\cite{haldane1988model}. While QAHE has been proposed in plaquette-based systems with ferromagnetism and SOC~\cite{Ye2015}, experimental realizations remain limited. Magnetic square–octagon materials, such as ferromagnetic $\alpha$-MnO$_2$, provide a promising platform. In the model, the staggered flux acts as an effective Peierls phase in the hopping amplitudes. These complex hoppings break time-reversal symmetry while preserving zero net flux, generating Berry curvature and enabling a quantum anomalous Hall (QAH) phase. Such flux patterns can emerge from loop-current states~\cite{bounoua2020loop} or orbital ordering, as in Kagome Cs$_2$LiMn$_3$F$_{12}$, where magnetism and SOC produce a QAH state~\cite{Ye2015}.

Figure~\ref{fig:S5} shows the band structure, Berry curvature, and anomalous Hall conductivity (AHC) of ferromagnetic $\alpha$-MnO$_2$~\cite{crespo2013}. The Berry curvature is concentrated near the valence band and gives a finite anomalous Hall conductivity (Fig.~\ref{fig:S5}(c)). Ferromagnetic ordering splits the spin bands, while SOC and the square–octagon geometry generate nonzero Berry curvature. This demonstrates that magnetism, lattice symmetry, and SOC together reproduce QAHE, consistent with our minimal staggered-flux model.

\subsection{Computational details}\label{computational}

First-principles calculations were performed within the DFT framework using projector augmented-wave (PAW) potentials, as implemented in the Vienna \textit{ab-initio} simulation package (VASP)~\cite{hohenberg1964,kreese1996}. Exchange-correlation effects were treated using the generalized gradient approximation (GGA), and SOC was included self-consistently to account for relativistic effects~\cite{perdew1996}. The in-plane lattice constants (here, $a=b$) are 3.44~\AA\ for Carbon, 6.33~\AA\ for MoS$_2$, 5.69~\AA\ for MoSi$_2$N$_4$, and 7.13~\AA\ for MoSi$_2$As$_4$. A plane-wave cutoff of 480~eV and Gaussian smearing with a width of 50~meV were applied to all compounds. Brillouin-zone sampling employed a $\Gamma$-centered $9 \times 9 \times 1$ $k$-mesh with an energy-convergence threshold of 10$^{-6}$~eV. A vacuum of 12~\AA\ was introduced for all materials to eliminate spurious interactions between periodic images in the out-of-plane direction. For the non-magnetic phases, we did not use an on-site Coulomb interaction for the transition-metal atoms. However, the ferromagnetic state of the $\alpha$--MnO$_2$ phase (space group $I4/m$, No.~87) was modeled by including an on-site Coulomb interaction for the Mn $d$ electrons within the GGA+$U$ scheme using $U = 3$~eV, which yields an effective magnetic moment of $\sim 3.2\,\mu_B$ on the Mn atoms. The lattice parameters are $a=b=9.922$~\AA\ and $c=2.9291$~\AA.  The material-specific tight-binding Hamiltonians were constructed via the \textsc{VASP2WANNIER90} interface~\cite{mostofi2008wannier90}. The edge states for MoS$_2$ were obtained using the semi-infinite Green's function method~\cite{sancho1985highly}, and the AHC for $\alpha$-MnO$_2$ was evaluated using the linear-response Kubo formula as implemented in the \textsc{WannierTools} package~\cite{wu2018wanniertools5}.

\bibliography{ref_supp}